\documentstyle[aps,prl,preprint,tighten]{revtex}

\begin{document}
\preprint{McGill/94-57}
\draft
\title{Scaling with no phase transition}
\author{David Seibert\thanks{Electronic mail (internet):
seibert@hep.physics.mcgill.ca.}}
\address{Physics Department, McGill University, Montr\'eal, P.Q.,
H3A 2T8, Canada}
\date{20 December 1994}
\maketitle
\begin{abstract}
The observation of scaling relations in ultra-relativistic nuclear
collisions would not by itself signal that the hot matter produced
in these collisions has passed through a phase transition.
\end{abstract}

One of the goals of studying the hot matter produced in
ultra-relativistic heavy ion collisions is to find a signal that the
produced matter is a quark gluon plasma, or some other exotic phase
of strongly-interacting matter, which is separated from low-temperature
hot nuclear matter by a phase transition.  One technique which has been
explored is the construction of multi-particle correlation functions
such as scaled factorial moments (SFMs)\cite{BP},
\begin{equation}
F_q = \frac {\langle n!/(n-q)! \rangle} {\langle n \rangle^q},
\end{equation}
where $n$ is the number of particles in a given bin.  The hope is that,
because fluctuations typically have non-trivial behavior near phase
transitions, the SFMs may also exhibit anomalous behavior near a phase
transition.

Recently, Mohanty and Kataria\cite{MK} showed that, if the hot matter
produced in an ultra-relativistic nuclear collision freezes out shortly
after passing through a first- or second-order phase transition, the
SFM data will obey a scaling law:
\begin{equation}
F_q \propto F_2^{(q-1)^{1.33}}.
\end{equation}
They then claimed that the observation of this scaling would signal that
the hot matter had frozen out shortly after passing through a phase
transition.  This scaling was previously demonstrated (for the case of a
second-order phase transition only) by Hwa and collaborators\cite{Hwa},
who claimed that it would indicate passage through a second-order phase
transition.  However, neither claim is true, as the scaling can be
produced even if the hot matter never passes through a phase transition.

Both groups worked in the context of a Ginzburg-Landau mean field theory
of phase transitions\cite{GL}.  In their calculations, the free energy
density of the hot matter is
\begin{equation}
F[\phi] = a'(T-T_c) \left| \phi \right|^2 + b \left| \phi \right|^4
+ c \left| \phi \right|^6,
\end{equation}
where $\phi$ is the order parameter for the phase transition.  The phase
transition is second-order for $b>0$ and first-order for $b<0$.

Mohanty and Kataria demonstrated that, within this Ginzburg-Landau mean
field theory, scaling of SFM data will occur if the hot matter freezes
out shortly after passing through a phase transition, whether first- or
second-order.  However, they neglected to show that it is necessary for
the matter to pass through the phase transition for the scaling to occur.
For example, the scaling may be observed if there is a phase transition
at finite temperature $T_c$ but infinite (or very high) energy density,
so that the system approaches $T_c$ from below but never crosses into
the high-temperature phase.  Scaling may also be observed if the equation
of state of the hot matter exhibits a fast but smooth transition, where
the coefficient of the quadratic term becomes small but never changes
sign so that there is no phase transition.  Thus, the mere observation of
scaling does not indicate that the hot matter has passed through a phase
transition.

\acknowledgements

This work was supported in part by the Natural Sciences and Engineering
Research Council of Canada, and in part by the FCAR fund of the Qu\'ebec
government.

\vfill \eject

\end{document}